\documentclass[superscriptaddress,reprint]{revtex4-2}  %two column

\usepackage{xcolor}
\usepackage{amsmath}    % Fancy math fonts
\usepackage{graphicx}   % Include figure files
\usepackage{dcolumn}    % Align table columns on decimal point
\usepackage{bm}         % Bold math
\usepackage[caption=false]{subfig}
\newcommand{\xm}{\relax\ifmmode X_{\mathrm{max}} \else
  $X_{\mathrm{max}}$\fi}
\newcommand{\mxm}{\relax\ifmmode \left<X_{\mathrm{max}}\right> \else
  $\left<X_{\mathrm{max}}\right>$\fi}
\newcommand{\sxm}{\relax\ifmmode \sigma(X_{\mathrm{max}}) \else
  $\sigma(X_{\mathrm{max}})$\fi}
\newcommand{\nm}{\relax\ifmmode N_{\mathrm{max}} \else
  $N_{\mathrm{max}}$\fi}

\begin{document}
\title{Measurement of the Proton-Air Cross Section with Telescope Array's Black Rock Mesa and Long Ridge Fluorescence Detectors, and Surface Array in Hybrid Mode}
\author{R.U. Abbasi}
\email{rabbasi@luc.edu}
\affiliation{Department of Physics, Loyola University Chicago, Chicago, Illinois, USA}

\author{M. Abe}
\affiliation{The Graduate School of Science and Engineering, Saitama University, Saitama, Saitama, Japan}

\author{T. Abu-Zayyad}
\affiliation{High Energy Astrophysics Institute and Department of Physics and Astronomy, University of Utah, Salt Lake City, Utah, USA}

\author{M. Allen}
\affiliation{High Energy Astrophysics Institute and Department of Physics and Astronomy, University of Utah, Salt Lake City, Utah, USA}

\author{R. Azuma}
\affiliation{Graduate School of Science and Engineering, Tokyo Institute of Technology, Meguro, Tokyo, Japan}

\author{E. Barcikowski}
\affiliation{High Energy Astrophysics Institute and Department of Physics and Astronomy, University of Utah, Salt Lake City, Utah, USA}

\author{J.W. Belz}
\affiliation{High Energy Astrophysics Institute and Department of Physics and Astronomy, University of Utah, Salt Lake City, Utah, USA}

\author{D.R. Bergman}
\affiliation{High Energy Astrophysics Institute and Department of Physics and Astronomy, University of Utah, Salt Lake City, Utah, USA}

\author{S.A. Blake}
\affiliation{High Energy Astrophysics Institute and Department of Physics and Astronomy, University of Utah, Salt Lake City, Utah, USA}

\author{R. Cady}
\affiliation{High Energy Astrophysics Institute and Department of Physics and Astronomy, University of Utah, Salt Lake City, Utah, USA}

% ...

\collaboration{The Telescope Array Collaboraton}
\noaffiliation


%\linenumbers\relax            % Comment out to disable line numbering
\begin{abstract}
  Ultra high energy cosmic rays provide the highest known energy
  source in the universe to measure proton cross sections. Though
  conditions for collecting such data are less controlled than an
  accelerator environment, current generation cosmic ray observatories
  have large enough exposures to collect significant statistics for a
  reliable measurement for energies above what can be attained in the
  lab. Cosmic ray measurements of cross section use atmospheric
  calorimetry to measure depth of air shower maximum (\xm), which is
  related to the primary particle's energy and mass. The tail of the
  \xm{} distribution is assumed to be dominated by showers generated
  by protons, allowing measurement of the inelastic proton-air cross
  section. In this work the proton-air inelastic cross section
  measurement, $\sigma^{\mathrm{inel}}_{\mathrm{p-air}}$, using data
  observed by Telescope Array's Black Rock Mesa and Long Ridge
  fluorescence detectors and surface detector array in hybrid mode is
  presented.  $\sigma^{\mathrm{inel}}_{\mathrm{p-air}}$ is observed to
  be $520.1 \pm 35.8$
  [Stat.] $^{+25.0}_{-40}$[Sys.]~mb at $\sqrt{s} =  73$ TeV. 
 The total proton-proton cross section is subsequently
  inferred from Glauber formalism and is found to be
  $\sigma^{\mathrm{tot}}_{\mathrm{pp}} = 139.4 ^{+23.4}_{-21.3}$ [Stat.]$ ^{+15.0}_{-24.0}$[Sys.]~mb.
\end{abstract}

 %rm for apj
\maketitle % rm for apj
\section{Introduction}\label{sec:intro}

Ultra high energy cosmic rays (UHECRs) offer a unique opportunity as
testing grounds for physics beyond the standard model, as they
represent a class of particles in the energy frontier beyond what can
be generated in human-made accelerators. In addition to questions
concerning their astrophysical nature, such as location of sources,
composition, acceleration mechanisms, and propagation modes,
fundamental aspects regarding the nature of matter can be investigated
as well. In particular, UHECRs provide a way to measure the proton
interaction cross section at energies beyond what can be achieved in
the lab to test standard model predictions of how the cross
section evolves with energy beyond what is measured in
accelerators. Whereas accelerators are highly controlled environments,
specially designed to maximize integrated luminosity, cosmic ray
experiments must rely on natural accelerators in the universe which
can not be tuned to deliver a desired luminosity. The only choice for
UHECR detectors is to increase their aperture to collect more events
given a fixed interval of collection time.

UHECR detectors do not directly observe the primary particle of
interest due to the extremely low flux of the spectrum ($\sim
10^{-30}$~eV$^{-1}$ m$^{-2}$ sr$^{-1}$
s$^{-1}$)~\cite{2019ICRC...36..298I}. Instead, the primary particle
enters the Earth's atmosphere and quickly interacts with an air
molecule generating an extended air shower which generates copious
amounts of fluorescence light with secondary particles reaching the
ground. Telescope Array collects $\sim 3000$ events per year with
energies $> 1$~EeV ($\sqrt{s} > 43$~TeV) with the surface detector
array~\cite{AbuZayyad:2012ru}, which runs continuously day and night
(100\% duty cycle), and $\sim 700$ events per year per each monocular
fluorescence detector station~\cite{Abu-Zayyad:2013jra}, which only
run on clear, moonless nights ($\sim 10\%$ efficiency), in the same
energy range.

\begin{figure}
  \centering
  \includegraphics[clip,width=0.95\columnwidth]{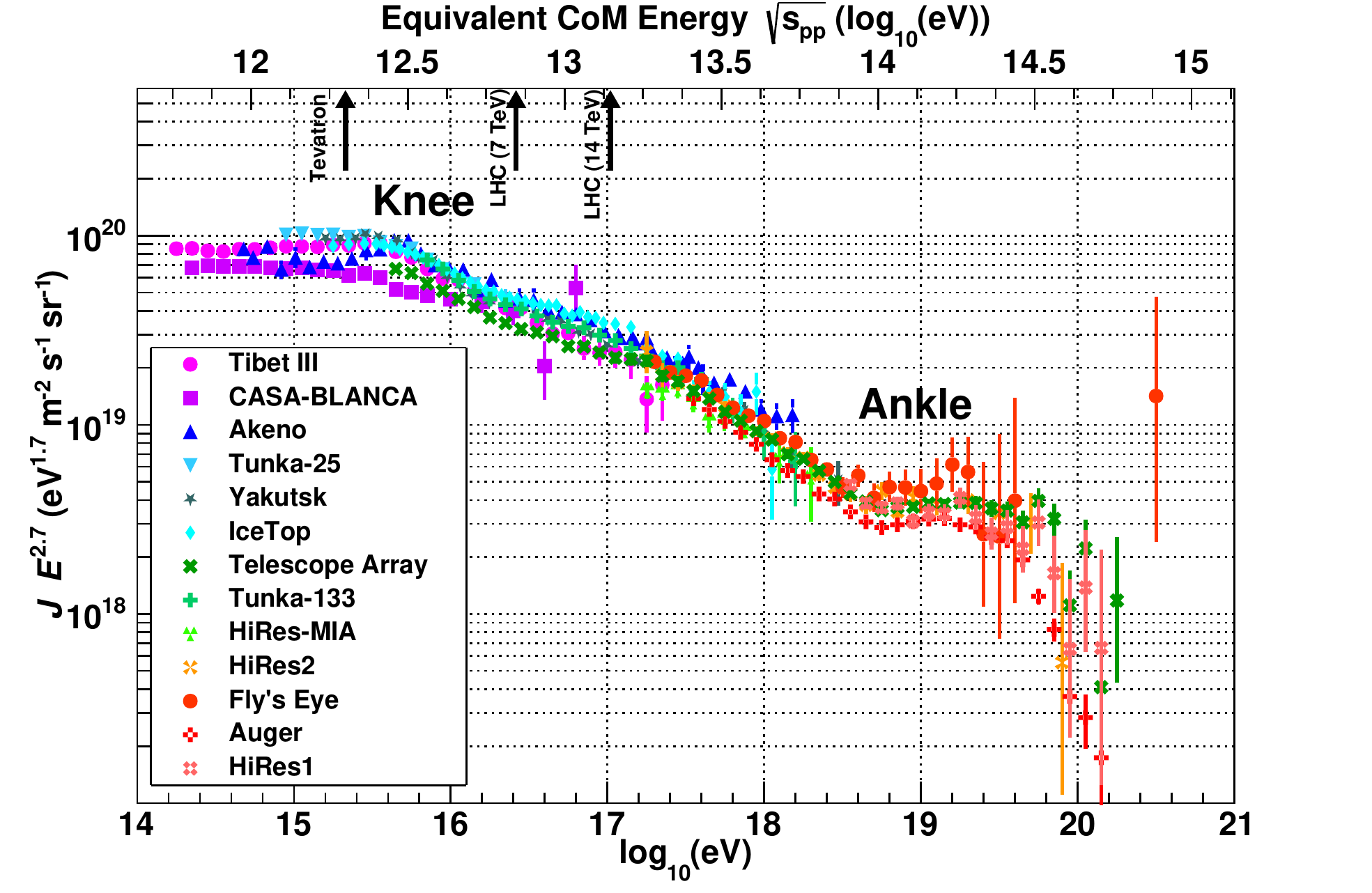}
  \caption{The cosmic ray spectrum starting at the knee, observed by
    recent experiments. Cosmic ray energies are measured in
    the lab frame (i.e., as a fixed target measurement), while the
    highest energy accelerator based measurements are done using
    colliding beams, reported in the center of momentum (CoM)
    frame. Recent accelerator experiments such as the Tevatron and LHC
    are closing the energy gap between human built accelerators and
    astrophysical accelerators. Data from
    \cite{Aartsen:2013wda,Abbasi:2007sv,AbuZayyad:2000ay,Amenomori:2008sw,Bird:1994wp,Fenu:2017hlc,Fowler:2000si,Ivanov:2015pqx,Knurenko:2015uoa,Nagano:1991jz,Prosin:2015voa}.}
  \label{fig:crspec}
\end{figure}

In an accelerator experiment cross sections are measured through
careful design and control of the source and target, using either
colliding beams or fixed target setup. Cross section, $\sigma$, in
colliding beams is determined by understanding the acceptance of the
detector and measuring the event rate for a given beam luminosity, $L
= \sigma^{-1} dN/dt$. A cosmic ray measurement of cross section is
more akin to a fixed target calorimeter, with the cosmic ray flux
acting as the beam and the atmosphere the target material. In the case
of UHECRs, the measurement of cross section is made in the lab frame
and the atmosphere can be treated as a fixed target since an incoming
proton has Lorentz factor $\gamma$ in excess of $10^9$ for $E \geq
1$~EeV. Figure~\ref{fig:crspec} shows the cosmic ray spectrum measured
over many decades of energy from the knee to the highest energies
observed. The top axis shows the equivalent center of momentum energy
of a proton-proton collision of the highest energy terrestrial
accelerators. UHECR energies are typically considered as events with
$E \gtrsim 1$~EeV ($\sqrt{s} \gtrsim 43$~TeV). As accelerator designs
are improved over time, human-made accelerators are closing the energy
gap between center of mass energies than can be achieved in the lab
and what is provided by nature.

\begin{figure}[!h]
  \centering
  \includegraphics[width=0.45\textwidth]{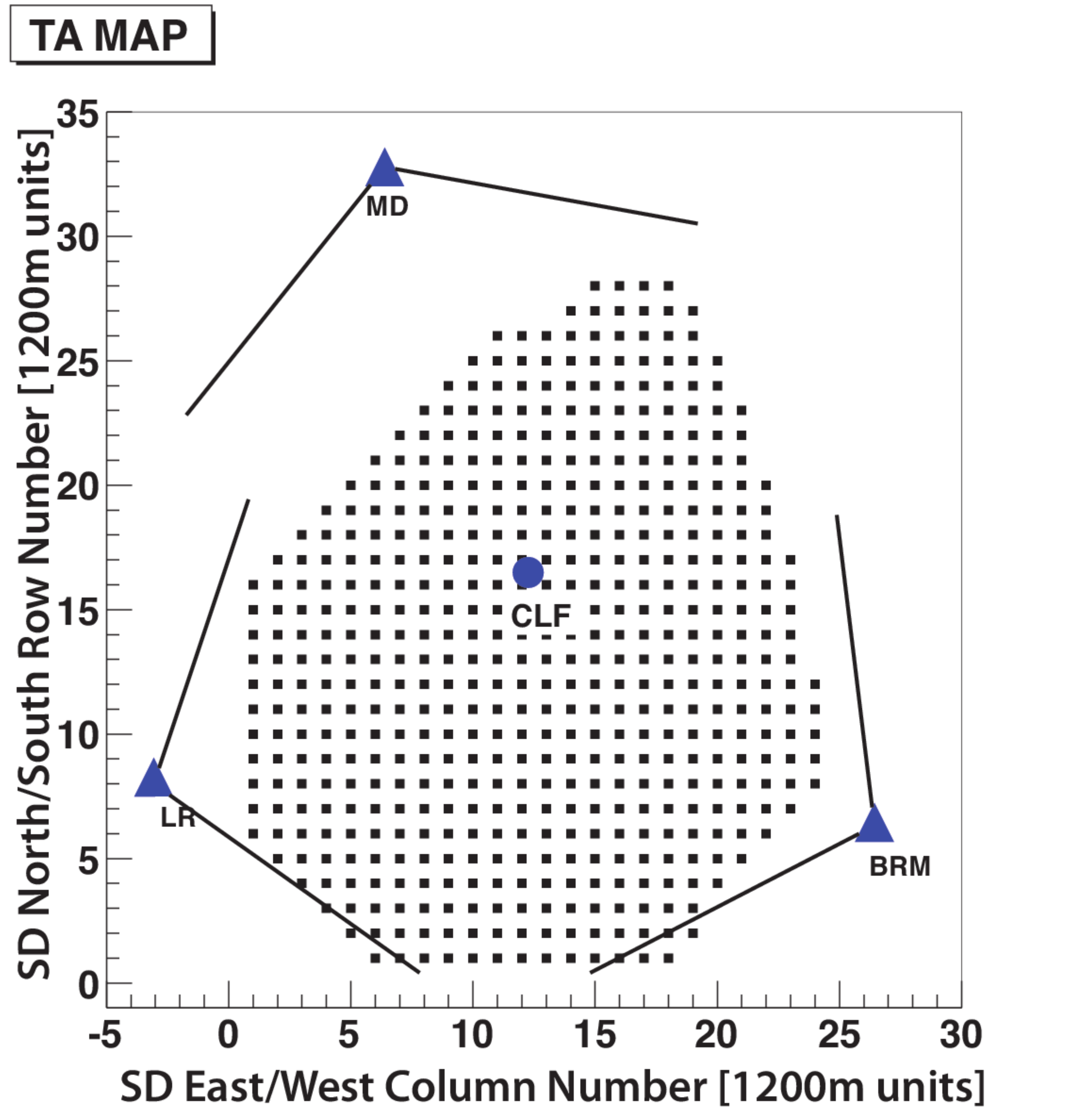}
    \caption{ The Telescope Array detector configuration. The filled squares are the 507 SD scintillators  on a~1.2 km grid. The SD scintillators are enclosed by three fluorescent detectors shown in filled triangles together with their field of view in solid lines. The northernmost  fluorescence detector is called Middle Drum while the southern fluorescence detectors are referred to as Black Rock Mesa and  Long Ridge. The filled circle in the middle equally spaced from the three fluorescence detectors is the Central Laser Facility used for atmospheric monitoring and detector calibration. }
    \label{fig:scheme}
\end{figure}

Ultra High Energy Cosmic Ray detectors have been reporting on the
proton-air cross section measurement beyond the capability of particle
accelerators since 1984~\cite{Siohan:1978zk, FE1987, Honda1992,
  Mielke:1994un, Belov2006, Aielli:2009ca, Auger2012, EAStop}. This work presents the second Telescope Array report on the proton-air
cross section. The first result was reported in 2015 using the Middle
Drum (MD) fluorescence detector and the surface detector in hybrid
mode~\cite{abbasi2015}. In this paper, we are reporting on the
inelastic proton-air cross section, at $\sqrt{s}=73$~TeV, using nearly
nine years of data observed by Black Rock Mesa (BRM) and Long Ridge (LR) 
fluorescence detectors (FDs) and the surface detector (SD) array in
hybrid mode. Note that the BRM and LR  detectors used in
this analysis, are closer in distance than MD to the surface detector array as
shown in Figure~\ref{fig:scheme}. This enables us to study the
inelastic proton-air cross section with higher statistical power for
lower energy events. The technique used to analyse these events is
similar to that used in the first proton-air cross section
report~\cite{abbasi2015}. The statistical power, on other hand,
increased by a factor of four. Note that all the systematic sources are
revisited and updated in addition to using the most recent hadronic
high energy models.

The proton-proton cross section is also calculated in this work using Glauber formalism~\cite{Glauber70} and BHS fit~\cite{Block05}. The inelastic proton-air and the total proton-proton cross section are compared to previous cosmic ray experimental results and to predictions from models.

%%%%%%%%%%%%%%%%%%%%%%%%%%%%%%%%%%%%%%%%%%%%%%%%%%
%\input{apparatus}
\section{Detector Description}\label{sec:apparatus}
Telescope Array (TA) is a cosmic ray observatory that deploys multiple
types of detectors to record the passage of extended air showers
caused by ultra high energy cosmic ray primaries as they
impact in the Earth's atmosphere. The primary way TA observes air
showers is by using surface detectors which detect the energy
deposited by high energy particles as they pass through them or using
fluorescence detectors which observe the UV light generated in
the atmosphere as the shower particles interact and exchange energy with
air molecules. SDs do not measure the development of the air shower in
the sky, while FDs do. The shower size, number of charged particles at
depth $X$ ($N(X)$), of an air shower can be parameterized using the Gaisser-Hillas
function~\cite{1977ICRC....8..353G}
\begin{equation}
  N(X) = \nm \left( \frac{X - X_0}{\xm - X_0} \right)^{\frac{\xm -
      X_0}{\lambda}} \exp\left( \frac{\xm - X}{\lambda}\right)
  \label{eq:gaisser-hillas}
\end{equation}
where $N$ is the number of particles at slant depth $X$. The
parameters \nm, \xm, $X_0$, and $\lambda$ describe the shower shape.
\nm{} is the maximum number of shower particles and the slant
depth at which this occurs is denoted by \xm{}. $\lambda$ and $X_0$
correspond to the proton-air interaction length and slant
depth of first interaction, respectively.

In practice, when fitting real shower profiles using the
Gaisser-Hillas function, $\lambda$ and $X_0$ are fixed parameters,
while \xm{} is observed by the FDs. To get an accurate measure of
\xm{}, a monocular FD measurement is not sufficient. To improve \xm{}
resolution, simultaneous observation of a shower by multiple FD
stations must be employed or simultaneous observation by a FD station
and the SD ground array. In the case of multiple FD stations, the
independently measured shower-detector planes provide a strong
constraint on the shower track, leading to greatly improved
geometrical resolution. Similarly, a shower observed by a single FD
station, along with the arrival time and core location on the ground
provided by the SD array delivers the same benefit. Resolution on
\xm{} improves dramatically from 84 and 52g/cm$^2$~g/cm$^2$ for showers
with energies of 1~EeV to 100~EeV
respectively~\cite{AbuZayyad:2000vea} to $\sim$20~g/cm$^2$ for $E >
1$~EeV when using multiple sets of observing stations to record
showers.

Telescope Array is located in central Utah's Millard County, USA. The SD ground array is composed of 507 plastic scintillator
counters spread over 700~km$^2$. The center of the SD array is located
at $39^\circ$ $17'$ $49''$N $112^\circ$ $54'$ $31''$W, 1370~masl. Three FD
stations overlooking the SD array are located outside the array
boundary. All FD stations are located $\sim$21~km away from the center of the
SD array. Middle Drum
station is located on the north end of the array, Black Rock Mesa is
located at the southeast border and Long Ridge is located at the
southwest border. This work uses FD data from the Black Rock Mesa and
Long Ridge detectors. While the general operation of all FDs is
similar, the design and location of the Middle Drum detector relative
to the SD array border results in different low energy acceptance than
the Black Rock Mesa and Long Ridge detectors. The description of FD
equipment used in this analysis that follows is for Black Rock Mesa
and Long Ridge.

Each FD station is comprised of twelve telescopes consisting of a
multi-segmented 6.8~m$^2$ mirror, a 16x16 photomultiplier tube (PMT)
array camera, electronics to digitize PMT signals at 40~MHz, trigger
on air shower track candidates, and readout and communications with a
remote DAQ which controls event readout and storage among all of the
electronics racks. The telescopes are arranged in a two ring
configuration providing zenith angle coverage in two bands. Six
telescopes are assigned to ring 1, observing $3^\circ - 17^\circ$ in
elevation angle and six are assigned to ring 2 observing $17^\circ -
31^\circ$. Azimuthal coverage of $108^\circ$ is the same for both
rings. On clear, moonless nights the FDs scan the skies for potential
air shower events. Because of this constraint, FD collection
efficiency is about 10\%, whereas properly operating SDs have 100\%
operating efficiency since SDs can operate in all weather conditions
24 hours a day. Each FD electronics rack has a track finder module
which implements temporal-spatial pattern recognition algorithms to
determine if a track has been observed. If a set of tube triggers
meets the criteria an event level trigger is generated and
communicated to the remote DAQ which forces readout of all mirrors for
storage and offline analysis. BRM and LR electronics utilize FADC
electronics which allows digitization of PMT signals at an equivalent
14-bit, 10~MHz sampling rate, allowing observation of the time
development of an event with 100~ns time resolution. Above
$10^{18.2}$~eV, showers are seen with distance of closest approach
(impact parameter) $> 25$~km. Further details about the construction
and design of the BR and LR stations can be found in
\cite{Tokuno,Tameda:2009zza}.

Each surface detector is composed of two layers of 3~m$^2$ plastic
scintillator, 1.2~cm thick. Wavelength shifting fibers are embedded in
grooves in the scintillator layers and optically coupled to a PMT (one
for each layer). PMT signals are digitized by a 12 bit FADC operating at
50~MHz sampling rate. Onboard electronics deployed with each SD scan
for signals above threshold ($> 3$~minimum ionizing particles) and
generate a trigger for signals that exceed this level. These triggers
are relayed to one of three remote DAQ stations by wireless radio
communications. The remote DAQ stations are responsible for generating
event level triggers based upon simple temporal-spatial pattern
matching. When a sufficient number of SDs submit triggers that meet
the criteria for an event level trigger, the remote DAQ station
broadcasts a directive for all SDs that observed signal above
a threshold to readout ($> 0.3$~minimum ionizing particles) and send their data to the DAQ for storage and
offline analysis. SDs are placed in a grid-like manner, with
separation distance of 1.2~km. SD array event reconstruction
efficiency saturates at $10^{18.9}$ and becomes 100\% efficient with
no zenith angle dependence. Refer to \cite{AbuZayyad:2012kk} for
further information regarding the technical details of TA's surface
detector array.

The second additional cut added to this analysis is a zenith angle
cut. Because the atmosphere acts as a calorimeter, fluorescence
detectors are limited in their ability to reconstruct showers based
upon the atmospheric overburden available for air showers to
develop. Fluorescence detectors far above sea level in general will
accept fewer showers for reconstruction than those closer to sea
level, because the higher detector has less atmosphere for the shower
to reach maximum size before hitting the ground. Thus, showers with small
zenith angle (close to vertical) traverse less atmosphere than those
with large zenith angle, reducing the chance that the shower will
achieve maximum size before penetrating the ground or falling below
the field of view of the detector.  Figure~\ref{fig:zenith_acceptance}
shows Telescope Array hybrid \xm{} acceptance of QGSJET~II.4
protons. As seen in the figure, events with zenith angle less than 30
degrees show a break in acceptance roughly corresponding to the
vertical depth of ground level, indicated by the dashed line showing
the vertical depth of the Central Laser Facility (CLF) at the center of the
SD array. Events that have zenith angle greater than 30 degrees, are
sufficiently inclined to provide enough slant depth to reach shower
maximum in the atmosphere. Indeed, these events show no significant
break in the \xm{} acceptance.

The proton-air cross section measurement uses information from the deep
tail of the \xm{} distribution under the assumption that only light
primaries such as protons, and possibly some helium contamination,
populate this region of the distribution. To ensure the highest level
of proton purity in the tail of the \xm{} distribution used to
determine $\sigma^{\mathrm{inel}}_{\mathrm{p-air}}$, we search for the
minimum zenith angle cut which results in nearly flat \xm{} acceptance for
all \xm{} in the energy range $18.2 \leq \log_{10}(E/\mathrm{ev}) <
19.0$. If \xm{} acceptance shows a break in the deep \xm{} region for
some range of zenith angles, those events must be removed because
showers induced by proton primaries may be lost in the \xm{}
distribution tail.

Analysis, data, and Monte Carlo used for this work is identical to
that used in \cite{Abbasi:2018nun} except for energy binning and the
additional zenith angle cut described above. The resultant data set
contains 1975 events with a resolution in \xm{} of $\sim$20~g/cm$^2$
and an average energy of $10^{18.45}$ eV. For further details
concerning the hybrid analysis procedure refer to
\cite{Abbasi:2018nun}.

%%%%%%%%%%%%%%%%%%%%%%%%%%%%%%%%%%%%%%%%%%%%%%
%\input{data}
\section{Data Trigger, Reconstruction, AND Selection.}\label{sec:data}

The FD and SD data streams are collected independent of each other. To
create a hybrid data stream the streams are searched for coincident
triggers that occur within 500 $\mu$s. For this set of hybrid events,
SD reconstruction proceeds as described in \cite{AbuZayyad:2012kk} to
determine the shower core location and arrival time. FD reconstruction
is performed as described in \cite{Abu-Zayyad:2013jra} to determine
the shower-detector plane for each FD station that observes a
shower. This determines the shower-detector plane angle, $\psi$,
impact parameter, core location, and arrival time. A hybrid
reconstruction takes the additional step of casting the individual SDs
into ``pixels'' that observe the shower in a similar way FD PMTs
do. This allows us to use them in the shower-detector plane
fit. Because of their accurate measure of the shower track position
and arrival time on the ground, these points provide an additional
constraint on the track geometry. Once the hybrid shower geometry is
determined, the shower profile is measured by each observing FD
station using this improved measure of the shower track. Shower
profile reconstruction determines the shower size measured by the
number of charged particles as a function of atmospheric depth
(equation~\ref{eq:gaisser-hillas}) and proceeds as described in
\cite{Abu-Zayyad:2013jra}. The shower profile is used to determine the
primary particle energy and \xm, both of which are the essential
inputs to the proton-air cross section measurement.

The data used for this analysis was collected from 2008 May 27 to 2016
November 29, nearly nine years, and is the same data used the BR/LR
hybrid \xm{} measurement in \cite{Abbasi:2018nun}. That analysis
examined \xm{} for events with $E >= 10^{18.2}$~eV and resulted in
3330 events after applying all quality cuts to the data. The present
analysis imposes two more cuts on the data required for a good quality
cross section measurement: here we restrict analysis to events with
energy $18.2 \leq \log_{10}(E/\mathrm{eV}) < 19.0$ and zenith angle $>
30^\circ$. The rational for these additional cuts is described below.

\begin{figure}
  \centering
  \subfloat[$18.2 \leq \log_{10}(E/\mathrm{eV})< 18.3$\label{fig:mxm_sxm_a}]{%
    \includegraphics[clip,width=0.8\columnwidth]{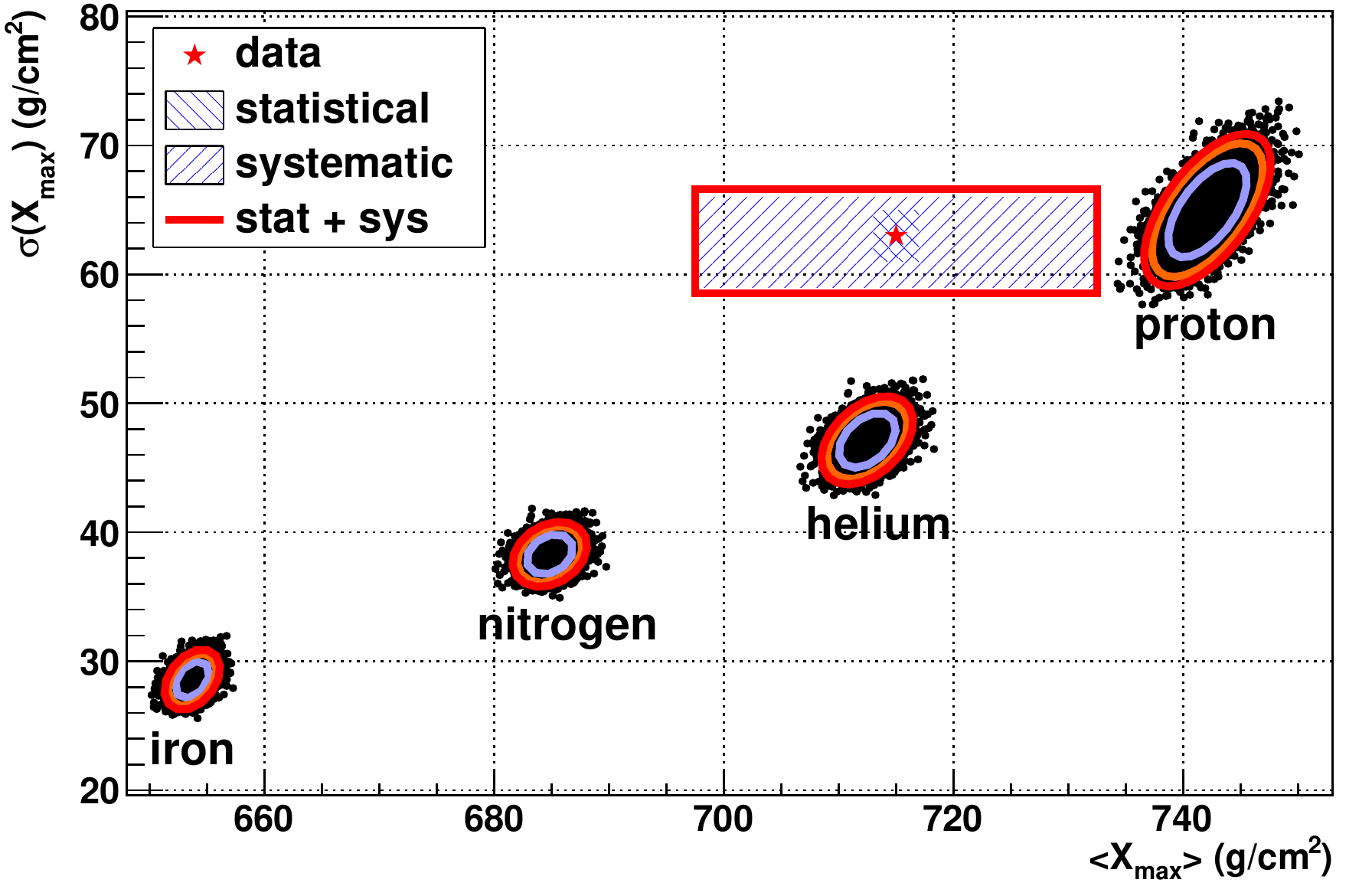}%
  }

  \subfloat[$19.4 \leq \log_{10}(E/\mathrm{eV}) < 19.9$\label{fig:mxm_sxm_b}]{%
    \includegraphics[clip,width=0.8\columnwidth]{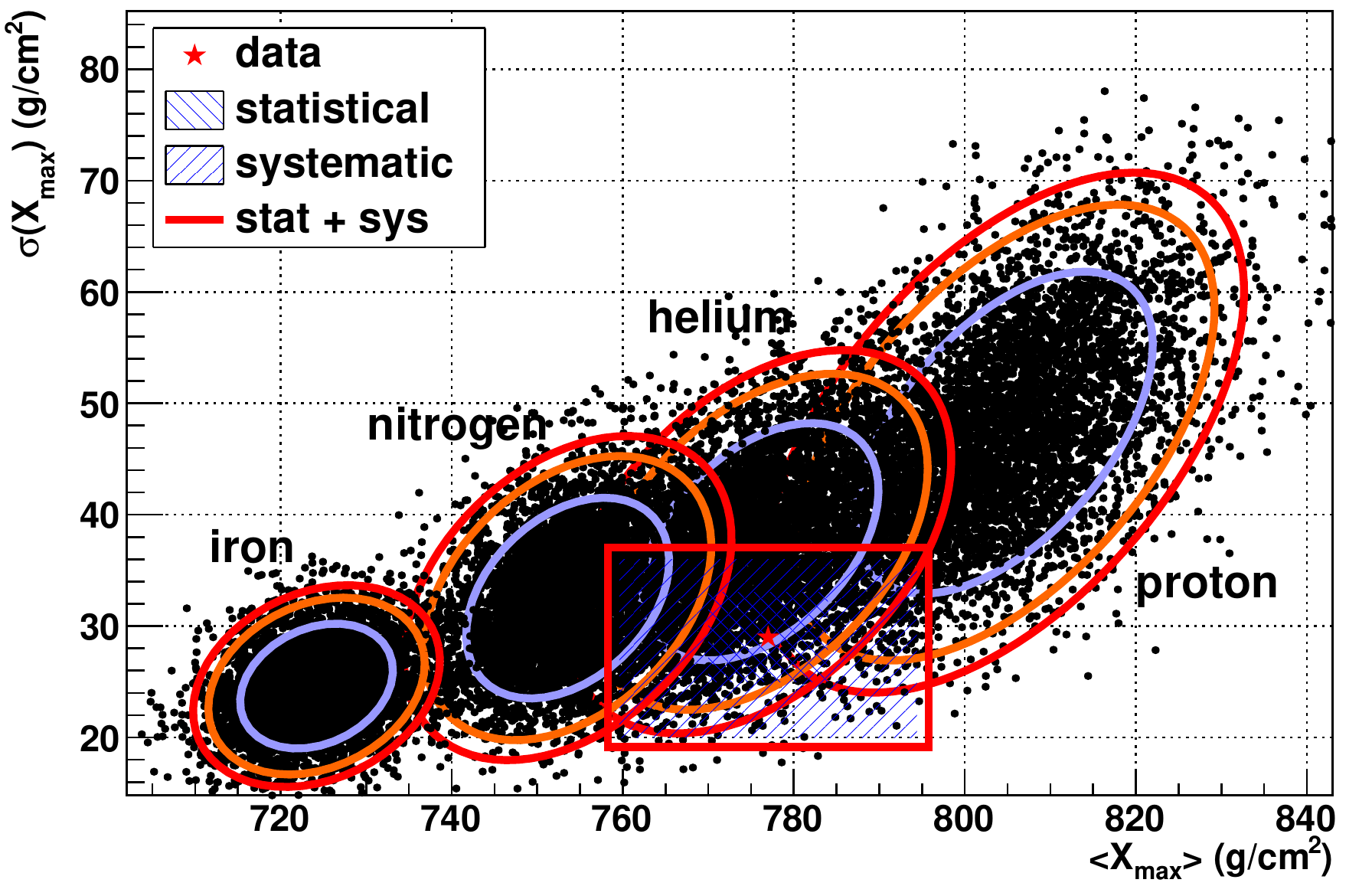}%
  }
  \caption{Measurements of data and QGSJet~II.4 Monte Carlo \mxm{}
    and \sxm{} in energy bins for $18.2 \leq \log_{10}(E/\mathrm{eV})
    < 18.3$ and $19.4 \leq \log_{10}(E/\mathrm{eV}) < 19.9$. The star
    represents \mxm{} and \sxm{} observed by TA in the two energy
    bins, as well as the statistical and systematic uncertainties.
    Each Monte Carlo chemical element shows the 68.3\% (blue ellipse),
    90\% (orange ellipse), and 95\% (red ellipse) confidence
    intervals.}
    \label{fig:mxm_sxm}
\end{figure}

\citet{Abbasi:2018nun} demonstrated that below $10^{19.0}$~eV, the TA RMS of the \xm{} distribution 
\sxm{} is consistent with light composition ranging between 52 and
63~g/cm$^2$. Above this energy \sxm{} begins to decrease. Due to changing
zenith angle acceptance and falling statistics it is premature to say if this
narrowing of \sxm{} is astrophysical in nature or caused by selection
bias. We can compare TA's observed mean \mxm{} and RMS \sxm of the
\xm{} distribution to Monte Carlo
predictions by randomly sampling the \xm{} distributions of individual
elements such as proton, helium, nitrogen, and iron according to data
statistics. To do this the simulated \xm{} distributions of each of
those elements is randomly sampled $N$ times, where $N$ is the number
of events observed in the data for the given energy bin, a
distribution of \xm{} is therefore generated, and \mxm{} and \sxm{} of
the distribution is recorded. Note the that \xm{} distributions are
fully simulated with all acceptance effects present in reconstructed
data. This procedure is repeated 5000 times. We can then measure the
68\%, 90\%, and 95\% confidence intervals of the joint expectation of
\mxm{} and \sxm{} for each element as shown. We can then compare the
predictions of \mxm{} and \sxm{} to what is observed in the data,
which is shown in figure~\ref{fig:mxm_sxm} for two energy bins. The
figure also shows \mxm{} and \sxm{} observed by TA, as well as the
systematic and statistical uncertainties.

Figure~\ref{fig:mxm_sxm_a} shows \mxm{} and \sxm{} observed for
$18.2 \leq \log_{10}(E/\mathrm{eV})< 18.3$ and the predictions for
primary particle spectra of pure proton, helium, nitrogen and iron
using the QGSJET~II.4 hadronic model. This is the lowest energy bin used
and the one with the most statistics (801 events) in that
analysis. \mxm{} and \sxm{} of the data are closest to the prediction
of QGSJET~II.4 protons. Additionally, the predictions from Monte Carlo
simulation have relatively small dispersion and are easily
distinguishable because of the relatively large statistics in this
energy bin. The TA hybrid data is tested against these
single element models and, in this energy bin, it is easy to see given
TA's statistical and systematic errors, as well as the clear
separation in simulated \mxm{} and \sxm{}, that the best fit to the
data is compatible to only one element within systematic errors. The
situation changes though where statistics are small, as shown in
figure~\ref{fig:mxm_sxm_b}. This energy bin shows \mxm{} and \sxm{}
observed for $19.4 \leq \log_{10}(E/\mathrm{eV})< 19.9$ as well as the
single element predictions. It only has 19 events and is the lowest
statistics bin in that analysis, with statistical power falling due to
the competing effects of a steeply falling primary particle spectrum
and zenith angle acceptance of hybrid reconstruction. Compared to
figure~\ref{fig:mxm_sxm_a}, observed \mxm{} here has increased as
predicted due to the relationship between primary particle energy and
\xm{} and observed \sxm{} has decreased. But the simulations predict
much larger dispersion in the \mxm{} and \sxm{}, causing the data to
be indistinguishable from light single element models such as proton,
all the way up to single element nitrogen. Above $E \geq
10^{19.0}$~eV the observed data exhibits this effect to such a degree
that we impose an additional cut eliminating events above this energy
for the present analysis to ensure that the tail of the \xm{}
distribution used for the p-air cross section measurement is not
significantly contaminated with heavy elements. We also estimate the
contamination in the tail due to helium and this estimate is described
later in this section.

\begin{figure}
  \centering
  \includegraphics[clip,width=0.9\columnwidth]{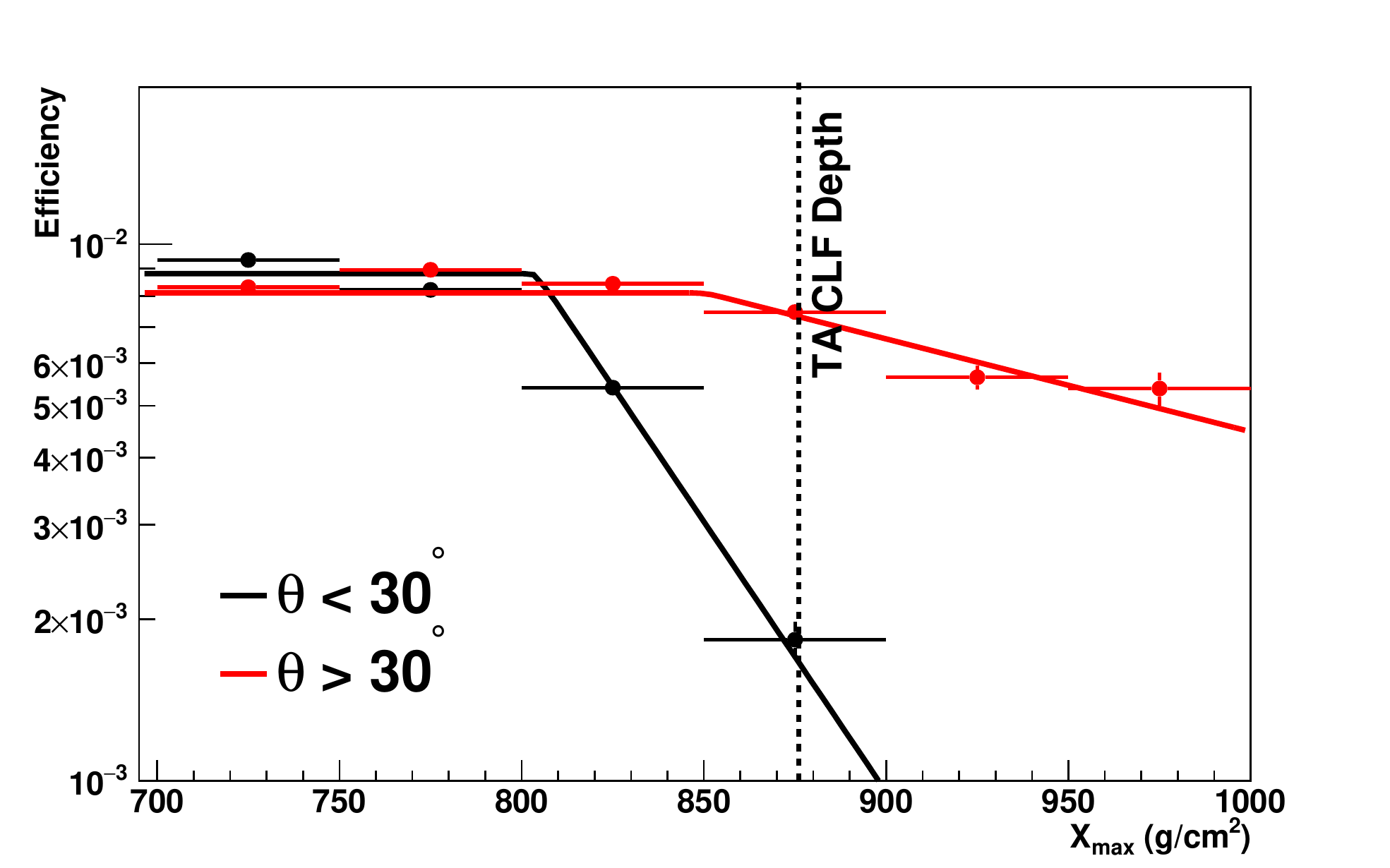}
  \caption{Telescope Array hybrid QGSJET~II.4 proton \xm{} acceptance
    in two zenith angle bands for energies $18.2 \leq
    \log_{10}(E/\mathrm{eV}) < 19.0$. The black shows the
    reconstruction efficiency of events with zenith angle $<
    30$~degrees and red line is the efficiency with zenith angle $>
    30$~degrees. Small zenith angle events are more likely to achieve
    shower maximum below the FD field of view and therefore the
    acceptance drops roughly at the vertical depth of ground level at
    TA. The dashed line shows the vertical depth of the TA Central
    Laser Facility (CLF). Inclined events with large zenith angle are more
    likely to reach shower maximum in the atmosphere and the
    acceptance does not significantly fall off with slant depth.}
  \label{fig:zenith_acceptance}
\end{figure}

The second additional cut added to this analysis is a zenith angle
cut. Because the atmosphere acts as a calorimeter, fluorescence
detectors are limited in their ability to reconstruct showers based
upon the atmospheric overburden available for air showers to
develop. Fluorescence detectors far above sea level in general will
accept fewer showers for reconstruction than those closer to sea
level, because the higher detector has less atmosphere for the shower
to reach maximum size before hitting the ground. Thus, showers with small
zenith angle (close to vertical) traverse less atmosphere than those
with large zenith angle, reducing the chance that the shower will
achieve maximum size before penetrating the ground or falling below
the field of view of the detector.  Figure~\ref{fig:zenith_acceptance}
shows Telescope Array hybrid \xm{} acceptance of QGSJET~II.4
protons. As seen in the figure, events with zenith angle less than 30
degrees show a break in acceptance roughly corresponding to the
vertical depth of ground level, indicated by the dashed line showing
the vertical depth of the Central Laser Facility (CLF) at the center of the
SD array. Events that have zenith angle greater than 30 degrees, are
sufficiently inclined to provide enough slant depth to reach shower
maximum in the atmosphere. Indeed, these events show no significant
break in the \xm{} acceptance.

The proton-air cross section measurement uses information from the deep
tail of the \xm{} distribution under the assumption that only light
primaries such as protons, and possibly some helium contamination,
populate this region of the distribution. To ensure the highest level
of proton purity in the tail of the \xm{} distribution used to
determine $\sigma^{\mathrm{inel}}_{\mathrm{p-air}}$, we search for the
minimum zenith angle cut which results in nearly flat \xm{} acceptance for
all \xm{} in the energy range $18.2 \leq \log_{10}(E/\mathrm{ev}) <
19.0$. If \xm{} acceptance shows a break in the deep \xm{} region for
some range of zenith angles, those events must be removed because
showers induced by proton primaries may be lost in the \xm{}
distribution tail.

Analysis, data, and Monte Carlo used for this work is identical to
that used in \cite{Abbasi:2018nun} except for energy binning and the
additional zenith angle cut described above. The resultant data set
contains 1975 events with a resolution in \xm{} of $\sim$20~g/cm$^2$
and an average energy of $10^{18.45}$ eV. For further details
concerning the hybrid analysis procedure refer to
\cite{Abbasi:2018nun}.

%%%%%%%%%%%%%%%%%%%%%%%%%%%%%%%%%%%%%%%%%%%%%%
%\input{analysis}
\section{Analysis}\label{sec:analysis}

The proton-air inelastic cross section $\sigma$ is related to interaction length $\lambda$ (mean
free path) by
\begin{equation}
  \lambda = 1/(n\sigma)
  \label{eq:lambda}
\end{equation}
where $n$ is the target particle density. The probability of an
interaction in a slab of target material of thickness $dx$ is $P(x) =
(1/\lambda) dx$. Given a ``beam'' of cosmic rays, beam intensity, $I$,
decreases with increasing number of slabs traversed as $dI =
-I/\lambda~dx$, leading to the expression of beam intensity, $I(x) =
I_0 \exp(-x/\lambda)$, where $I_0$ is the initial intensity and $x$ is
depth. Therefore for cosmic rays the interaction length can be
measured by fitting a distribution of depth of first interaction
($X_0$) between the cosmic ray primary particle and an air nucleus to
find $\lambda$. In practice this is not feasible because the starting
point of the upper atmosphere is not well defined due to its very low
density and there is no appreciable fluorescence signal generated at
first interaction.

After the initial inelastic collision an air shower continues to grow
in size through production of secondaries mainly by radiative
processes of pair production and bremsstrahlung in the electromagnetic
portion of the shower. The shower grows until it reaches a maximum
size, \xm, dependent primarily on primary particle energy and mass,
then decreases as energy loss of secondaries becomes dominated by
collisional processes. \xm{} therefore is a uniquely defined point in
the shower profile that is observed by fluorescence detectors and can
be used as a proxy for $X_0$ to determine the interaction length of a
distribution of cosmic ray primaries. Proton-air cross section is therefore measured indirectly for air showers.

In this work, the \textit{K-Factor method}~\cite{abbasi2015} is used to obtain the proton-air cross section. The tail of the \xm{} distribution retains the exponentially falling
nature of the $X_0$ distribution encoded within it and can be
parameterized as $f(\xm) = \exp(-\xm/\Lambda_m$), where $\Lambda_m$ is
the exponential slope of the tail. The $K$-Factor method relates the
slope of the tail of the \xm{} distribution and to the slope of $X_0$
distribution, through a constant factor $K$ which is determined by MC simulation and is close to unity. 
\begin{equation}
  \Lambda_m = K \lambda_{\mathrm{p-air}}
  \label{eq:lambda_m_1}
\end{equation}
where we now label $\lambda_{\mathrm{p-air}}$ as the proton-air
interaction length and both $\Lambda_m$ and $\lambda_{\mathrm{p-air}}$
are measured in g/cm$^2$. Using the relationship between $\lambda$ and
$\sigma$ expressed in equation~\ref{eq:lambda}, cross section can be
related to the mean target mass of air as
\begin{equation}
\sigma^{\mathrm{inel}}_{\mathrm{p-air}} = \frac{\left<
  m_{\mathrm{air}} \right>}{\lambda_{\mathrm{p-air}}}
\label{eq:sig_pair1}
\end{equation}
and substituting this into equation~\ref{eq:lambda_m_1} we find
\begin{equation}
  \Lambda_m = K \frac{24160}{\sigma_{\mathrm{p-air}}} = K \frac{14.45
    m_{\mathrm{p}}}{\sigma_{\mathrm{p-air}}}
  \label{eq:lambda_m_2}
\end{equation}
where $\left< m_{\mathrm{air}} \right> = 24160$~mb~g~cm$^{-2}$ or $14.45
m_{\mathrm{p}}$ with the proton mass expressed in
g~\cite{Ulrich:2009zq}. $\sigma^{\mathrm{inel}}_{\mathrm{p-air}}$ is expressed in
mb and $\Lambda_m$ is in g/cm$^2$. This equation directly links
the observed \xm{} distribution to the proton-air cross section.

This $K$-Factor method is the same method used in the first TA report on the proton-air cross section in 2015~\cite{abbasi2015}. The data analysis here is divided into two parts. The first part is done by calculating the value of the  attenuation length ($\Lambda_{m}$) of the observed UHECR events.  In the second part, we calculate the inelastic proton-air cross section ($\sigma^{\rm inel}_{\rm p-air}$) value from the obtained attenuation length $\Lambda_{m}$. 

\subsection{Measuring the  Attenuation Length $\Lambda_{m}$}

The value of attenuation length $\Lambda_{m}$, and therefore the proton-air cross section, can be calculated by fitting the \xm{} distribution tail to the exponential function  $\exp(-\xm/\Lambda_m)$. Here only the tail of the \xm{} distribution is used to obtain $\Lambda_{m}$, because it is the most penetrating part of the distribution and is assumed to be composed mostly of protons. UHECR composition can not be measured on an
 event by event basis and must be inferred from a distribution of
 events. By restricting the determination of $\Lambda_m$ to the tail of
 the \xm{} distribution, potential contamination from heavier elements
 in the primary spectrum is reduced. 

The choice of the starting point of the tail fit ( the lower edge of the fit range) $X_i$ for the exponential fit is made by fitting the \xm{} distribution tail to two exponential functions with separate power indices. The break point of these two fits ( found to be at 790 g/cm$^2$ ) describe the best fit beyond which the distribution can be described using a single exponential function. This maximizes the number of events in the tail distribution while minimizing instability in the value of $\Lambda_{m}$ due to possible detector bias or helium contamination.  

\begin{figure}[ht]
\begin{center}\vspace{1cm}
\includegraphics[width=1.0\linewidth]{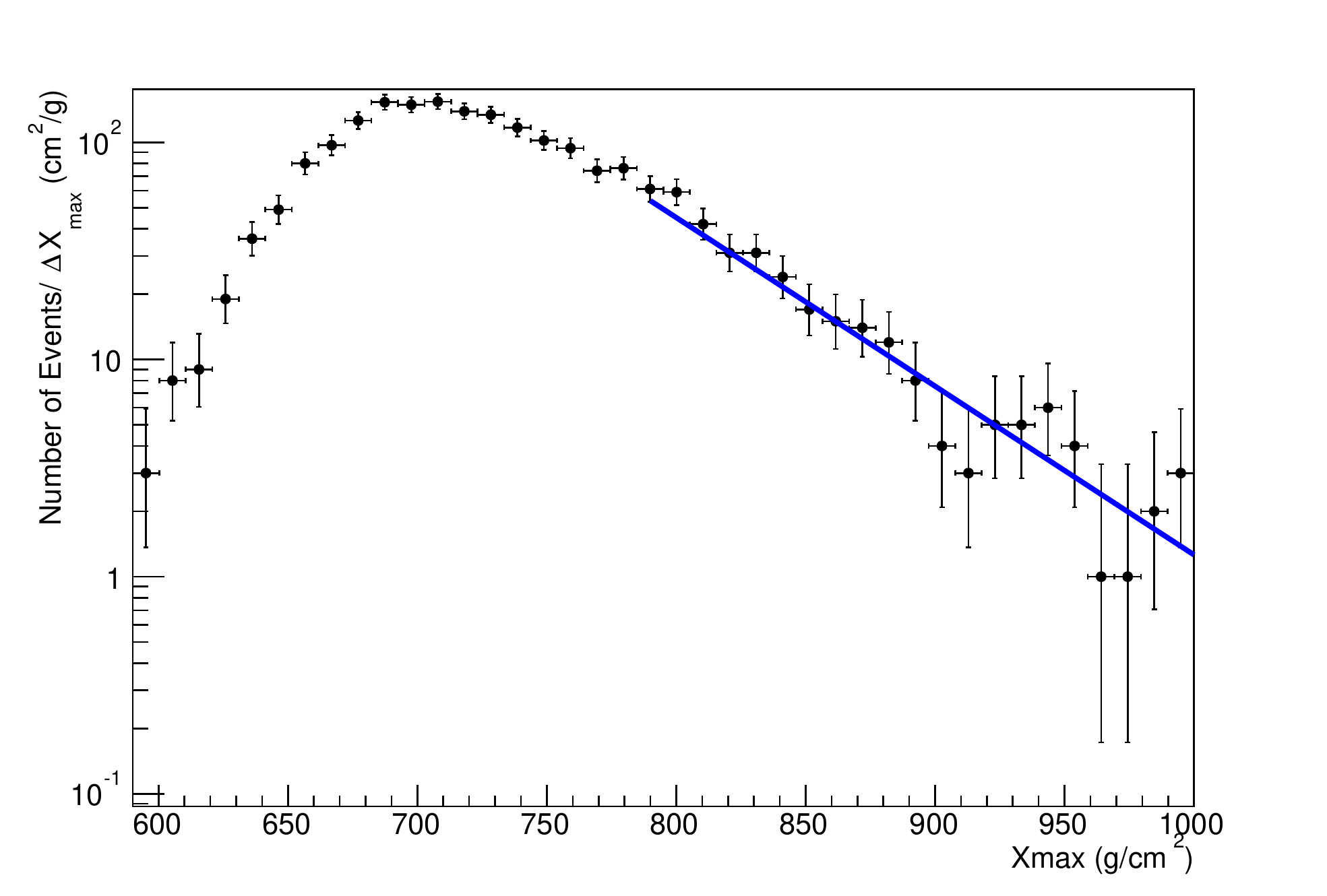}
\vspace{1.0cm}
\caption{\label{fig:xmax}  The number of events per \xm{} bin ($\Delta \xm$)  vs. \xm{} g/cm$^{2}$ for BRM and LR fluorescence detectors and the Telescope
      Array surface detector in hybrid mode. The line is the exponential fit to the slope using the unbined likelihood method between 790-1000 g/cm$^2$.  
}
\end{center} 
\vspace{1cm} 
\end{figure}

Figure~\ref{fig:xmax} shows the \xm{} distribution of the data collected by the Telescope Array southern most fluorescence detectors, Black Rock Mesa and Long Ridge, together with the surface detector hybrid events. The distribution includes 1975 events in the energy range between $10^{18.2}-10^{19.0}$ eV with an average energy of  $10^{18.45}$~eV. The data included here passed the quality cuts described in the previous section and is fitted to the exponential function  $\exp(-\xm/\Lambda_m)$ using the unbined maximum likelihood method.

Several systematic checks are applied to test for the  stability of the measured  attenuation length $\Lambda_{m}$. This is done by dividing the data in two halves based on: the  zenith angle, the distance of the shower using the impact parameter $R_p$, and the energy of the event. The divided subsets are found to be consistent within statistical fluctuations. 

The final $\Lambda_{m}$ measured by the Telescope Array detector at an average energy of $10^{18.45}$ eV ($\sqrt{s} = 73$~TeV) including the statistical checks is found to be $\Lambda_{m}$ = 55.9 $\pm$ 3.8 [Stat.] g/cm$^2$. Note that $\Lambda_{m}$ is directly derived from the data and is model independent.  Therefore, it can be used at a later time to calculate the inelastic proton-air cross section independent of the method or the UHECR models used in this paper.

\subsection{Proton-Air cross Section Measurement}

To determine the interaction mean free path of protons in air, $\lambda_{\mathrm{p-air}}$ and therefore the inelastic proton-air cross section $\sigma^{\mathrm{inel}}_{\mathrm{p-air}}$ we use the $K$-Factor technique. Using equation \ref{eq:lambda_m_1}, $K$ can be directly computed using Monte Carlo.  
$K$ depends on the hadronic model being used in
simulations. UHECR simulations rely on
the choice of electromagnetic
interaction driver, low energy hadronic generator, and high energy
hadronic generator~\cite{Heck:1998vt}. The most popular high energy
hadronic generators are SIBYLL2.3~\cite{sibyll,Ahn:2009wx},
EPOS-LHC~\cite{Pierog:2013ria},  QGSJET~II.4~\cite{Ostapchenko:2007qb,Ostapchenko:2010vb}, and  QGSJET01~\cite{qgsjet01}  which, with the exception of QGSJET01, are tuned to the most recent
accelerator data at energies accessible to accelerators and
extrapolated to UHECR energies through theoretical and
phenomenological predictions. Hadronic model dependence is an important and
difficult consideration when dealing with questions related to the
fundamental properties of hadronic air showers such as proton-air
cross section or cosmic ray composition. Some of the important
parameters that affect shower development which are extrapolated from
accelerator data to UHECR energies are inelasticity, multiplicity, and
cross section. Each hadronic generator uses different methods to do
this leading to differences in shower development at ultra high
energies. For a summary of these issues refer to
\cite{Engel:2011zzb,Ulrich2010}. For this work we present the results
for several different models and report on the systematic uncertainty
in the results of our measurement.

\begin{figure}
  %\begin{center}
  \centering
\includegraphics[width=1.0\linewidth]{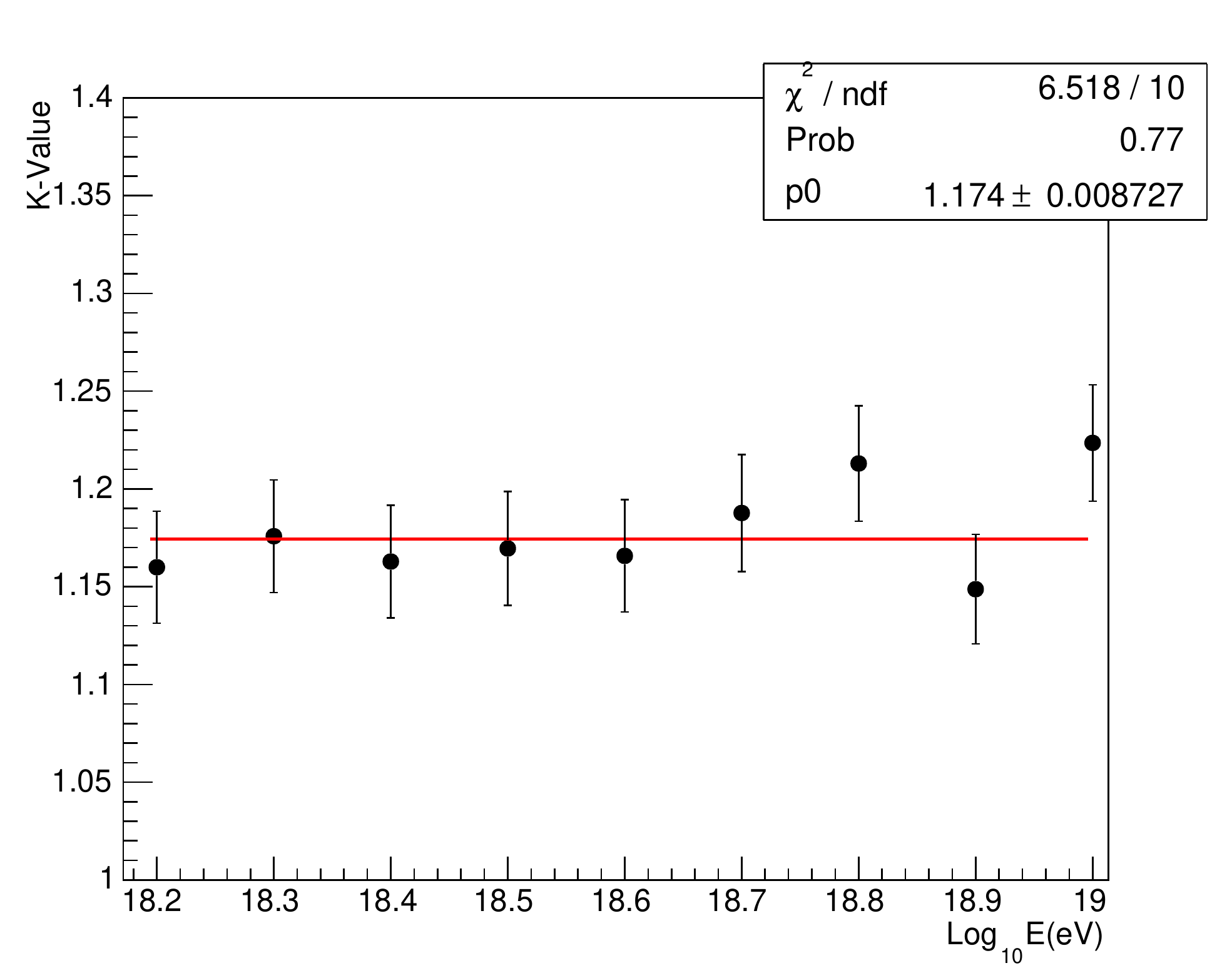}
%\vspace{1.0cm}
\caption{\label{fig:kval}  The value of $K$ obtained vs. energy in Log$_{10}$(eV)
for simulated data sets using CONEX 6.4 with the high energy
model QGSJET~II.4, for the energy range of the data, between
10$^{18.2}$ and 10$^{19.0}$ eV.}

%\end{center} 
%\vspace{1cm} 
\end{figure}

\begin{figure}[!h]
   \centering
  \includegraphics[width=1.0\linewidth]{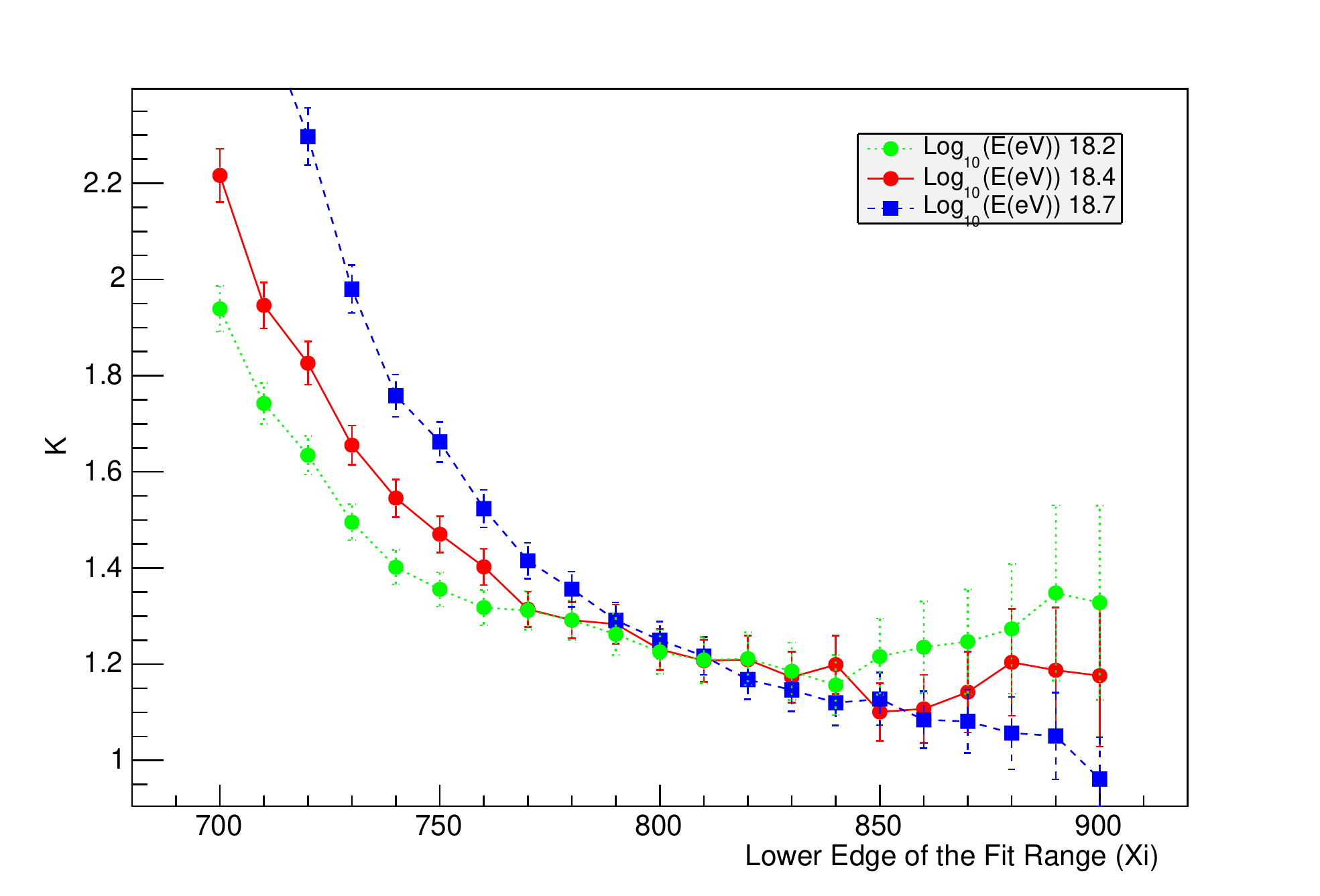}
     \caption{\label{fig:concor}  The value of $K$ vs. the lower edge in the fit  range ($X_i$) to the
      tail of  the \xm{} distribution for several data sets $10^{18.2}, 10^{18.4},$ and
      $10^{18.7}$ eV simulated using CONEX with the high energy
      model QGSJET~II.4. Each data set contains 10,000
      simulated events. 
     } 
     
 \end{figure}

$K$ is computed in this work by generating several simulated sets between $10^{18.2}-10^{19}$~eV for each of the high energy models. Each generated set contains ten thousand events using a one-dimensional air shower Monte Carlo program CONEX~6.4~\cite{conex,conex1,conex2}. Figure~\ref{fig:kval} shows the $K$-value including the statistical fluctuation calculated for each of these simulated sets, using QGSJET~II.4 as an example. The value of $K$  is then obtained by fitting  the $K$-value  vs. energy to a horizontal line as shown in Figure~\ref{fig:kval}.  It is important to note that the value of $\Lambda_{m}$  and therefore, the value of $K$ is dependent on the choice of the lower edge of the tail fit range $X_i$ (as shown in Figure~\ref{fig:concor} ).  A consistent procedure needs to be used to determine $X_i$  and therefore the value of $K$ for each energy bin and the high energy model shower simulations. To do so we calculate the difference in slant depth $D$ between the peak of the \xm{} distribution and 790~g/cm$^2$, using a simulated data set at an energy of $10^{18.45}$ eV (equivalent to the mean energy of the data set used in this work). The same difference in slant depth $D$ is later used to consistently determine the value of $X_i$ from peak of the \xm{} distribution for each of the simulated sets for each of the high energy models.

%\begin{figure}[ht]
%\begin{center}\vspace{1cm}

\begin{figure}[!h]
  \centering
  \includegraphics[width=1.0\linewidth]{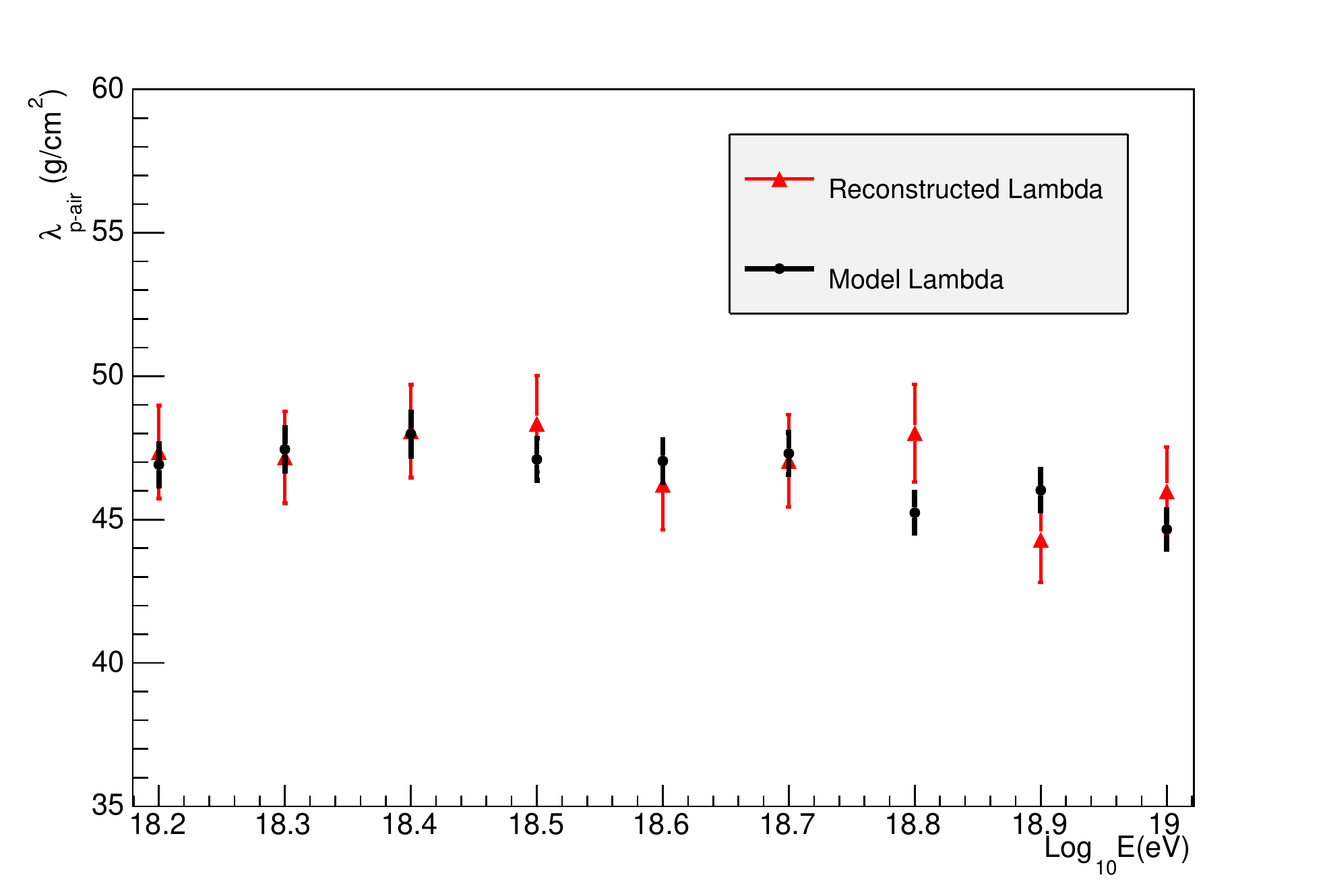}
    \caption{The proton-air interaction length $\lambda_{\rm p-air}$ in g/cm$^{2}$ vs. Energy in
      Log$_{10}$(eV) for the simulated data sets using CONEX with the high energy
      model QGSJET~II.4, for the energy range of the data, between $10^{18.2}$ and
      $10^{19.0}$ eV. The circle points are the   $\lambda_{\rm p-air}$
      values obtained from the $X_0$ distribution. Triangle points are the ones determined from reconstructing the
      $\lambda_{\rm p-air}$ values using the $K$-Factor method.}   
    \label{fig:lamlge}
\end{figure}

To confirm the validity of the obtained $K$ values, for each of the generated data sets, for each of the high energy models, $\lambda_{\mathrm{p-air}}$ is reconstructed and compared to the $\lambda_{\mathrm{p-air}}$ provided by the corresponding high energy model.  Figure~\ref{fig:lamlge} shows the comparison of the the values of the high energy model $\lambda_{\mathrm{p-air}}$ and the obtained $\lambda_{\mathrm{p-air}}$ using the $K$-Factor technique. Figure~\ref{fig:lamlge} shows that the value of $K$ obtained in this study indeed describes the value of $K$ of the high energy models correctly.

The $K$ value is dependent on the high energy model used.  The
obtained $K$ value is shown using CONEX 6.4 in
Table~\ref{tab:kval}, together with the corresponding inelastic proton-air cross
section $\sigma^{\mathrm{inel}}_{\mathrm{p-air}}$. Note that, the $\sigma^{\mathrm{inel}}_{\mathrm{p-air}}$ is calculated using equation~\ref{eq:lambda_m_2} with $\Lambda_{m}$ obtained from the TA \xm{} distribution and $K$ tabulated for each the high energy models QGSJET~II.4~\cite{Ostapchenko:2007qb,Ostapchenko:2010vb}, QGSJET01~\cite{qgsjet01} , SIBYLL2.3~\cite{sibyll,Ahn:2009wx}, and EPOS-LHC~\cite{Pierog:2013ria}.

Each $K$ listed in Table I is the average value of $K$ over the energy range of  $10^{18.2}$-$10^{19.0}$~eV. 
The value of $K$ is measured to be $\sim 20$\%
larger than 1.0 meaning the slope of the tail of the \xm{} distribution falls
more slowly than the $X_0$ tail. This is because the \xm{}
distribution resembles a convolution of a falling exponential, from
 the contribution of $X_0$, and a Gaussian from the growth of the
 shower and fluctuations of stochastic processes of shower
 development~\cite{Peixoto:2013tu}.  Showers exhibit large intrinsic
 fluctuations in development even for those initiated by particles of
 the same mass and energy. If showers did not exhibit these fluctuations, air shower \xm{} distributions would resemble the distribution of $X_0$, just
shifted to a greater depth in the atmosphere by a constant amount. 
It is important to note that the $K$ value model dependence shown in Table~\ref{tab:kval} is on the order of $\sim \pm$3$\%$ ($K$-Value historical improvement is discussed in~\cite{abbasi2015}). This makes the $K$-Value method  weakly model dependent and thus a reliable method to use in calculating the $\sigma^{\mathrm{inel}}_{\mathrm{p-air}}$.

\begin{center}
  \begin{table}   [!h]
    \begin{tabular}{| p{2.1cm} | p{2cm} | p{2cm} |}
      \hline
      Model & $K$ & $\sigma^{\rm inel}_{\rm p-air}$ (mb)\\ \hline
      QGSJET~II.4 &  $1.17 \pm 0.01$ & $505.4 \pm 34.8$\\ \hline
      QGSJET01   &  $1.19 \pm 0.01$ & $514.1 \pm 35.4$\\ \hline
      SIBYLL2.3     &  $1.24 \pm 0.01$ & $535.6 \pm 36.9$\\ \hline
      EPOS-LHC   &  $1.22 \pm 0.01$ & $527.0 \pm 36.3$\\
      \hline
  \end{tabular}
     \caption{ The value of $K$ obtained for each of the high energy
       models and the corresponding inelastic proton-air cross section for that model. Each $K$ listed is the single average value of $K$
       over the energy range of  $10^{18.2}$-$10^{19.0}$~eV. Note that the values of
       $K$ shows a~$\sim \pm$3$\%$ model uncertainty. }
   \label{tab:kval}%                                                            
\end{table}
\end{center}
%The average value of the inelastic proton-air cross section including the statistical fluctuation is found to be $\sigma^{\rm inel}_{\rm p-air} = 520.1 \pm 35.8$ [Stat.]~mb. 

In order to quantify the systematic uncertainties in the  $\sigma^{\rm inel}_{\rm p-air}$ measurement, several check were applied. First, systematic uncertainty due to model dependence is reported. This is done by quantifying 
the maximum variation in the $\sigma^{\rm inel}_{\rm p-air}$ value by each model from the average $\sigma^{\rm inel}_{\rm p-air}$ obtained from all of the high energy models. This uncertainty was found to be equal to $\pm 15$ mb. 

In addition, the systematic effect of possible energy dependent bias in the \xm{} distribution was studied. This was done by shifting the values of \xm{} by their elongation rate (the rate of change of shower \mxm{} w.r.t. shower energy) prior to fitting. The systematic effect from a possible energy bias was found to be negligible.

The systematic effects due to detector bias is also tested. This systematic effect is done by comparing the attenuation length calculated with and without detector effects. First, the attenuation length is calculated from simulations, where the detector effect is not included. After which, $\Lambda_{m}$ is calculated from simulation, where the events are propagated through the detector, reconstructed, and the quality cuts applied. 
The value of  $\Lambda_{m}$ was found to be consistent, for all the high energy models, between the thrown events and the reconstructed events with quality cuts applied. Therefore, the systematic effect from this test was found to be negligible. For further details concerning this test procedure refer to~\cite{abbasi2015}.

Another systematic check is done by studying the impact of contamination from other primaries. The 
systematic effect of other elements in the tail beside proton including  photon, CNO, helium and iron is investigated. Only photons and helium introduce a bias in the inelastic proton-air cross section. 

The upper limit of cosmic-ray photon fraction at the energy range in this study is found
to be $\sim$1.0$\%$, which is the best upper limit in the Northern hemisphere reported from the Yakutsk air-shower array~\cite{Glushkov:2009tn}.  The systematic uncertainty due to 1.0$\%$ gamma contamination is found to be $+20$~mb.  The contamination of helium in Telescope Array data between $10^{18.2}$-$10^{19.0}$~eV is measured to not larger than 43.8\% at the 95\% c.l. Using this limit, the systematic uncertainty due to helium contamination is found to be $-40$~mb. 

Note here that the sign for the systematic uncertainty due to helium
and gamma contamination is negative and positive respectively. Helium has a
larger cross-section than protons. Therefore, helium contamination will result
in the observation of a larger cross section than would be the case with pure
protons. The opposite occurs due to gamma contamination. The final systematic uncertainty for the $\sigma^{\mathrm{inel}}_{\mathrm{p-air}}$ is calculated by adding each of the systematic uncertainties quadratically.  

\begin{figure}[ht]
\begin{center}\vspace{1cm}
\includegraphics[width=1.0\linewidth]{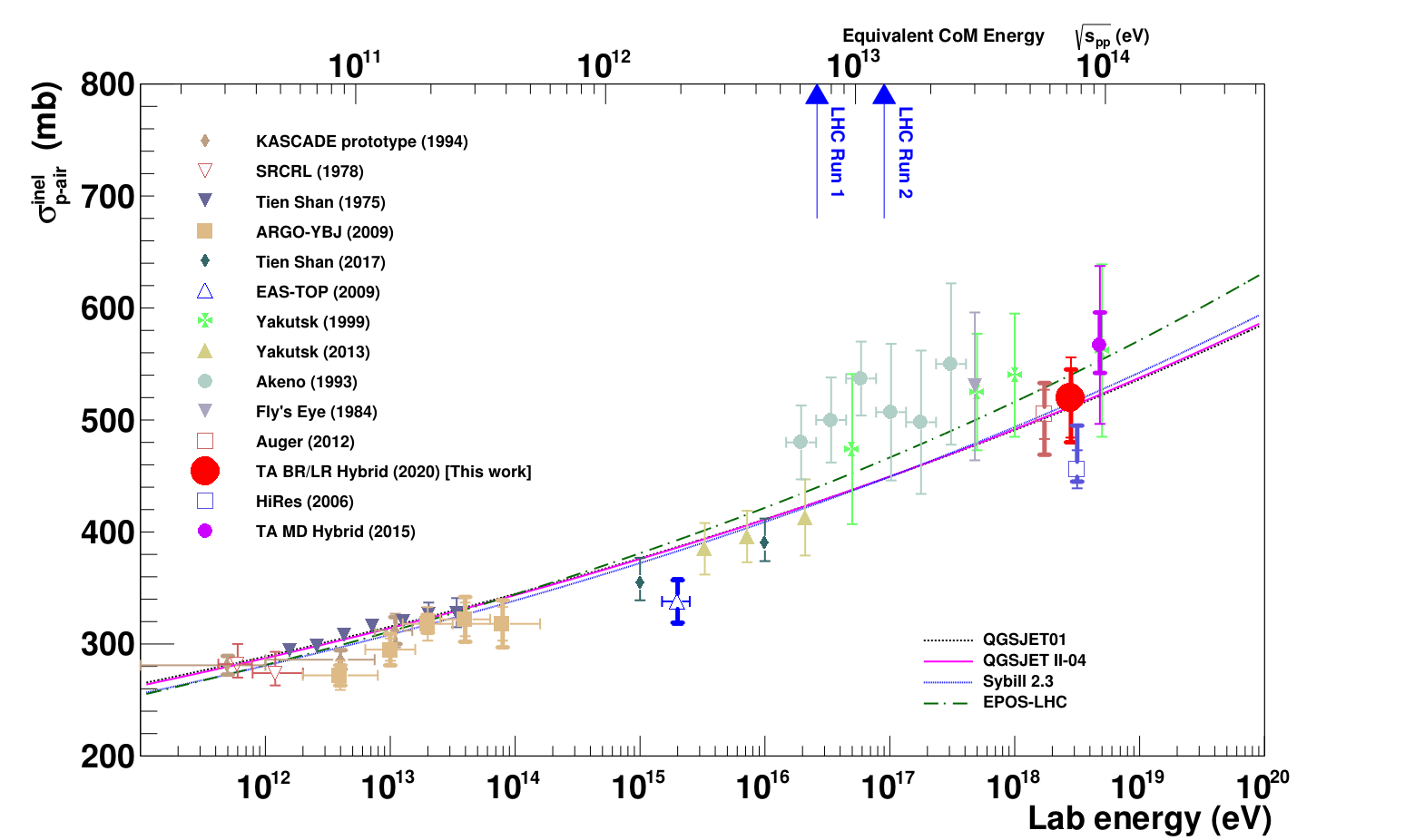}
\vspace{1.0cm}
\caption{\label{fig:pair}  The proton-air cross section result of this
  work, including the statistical (thin) and systematics (thick) error bars, in
      comparison to previous experimental results
~\cite{Siohan:1978zk, FE1987, Honda1992, Mielke:1994un, Belov2006, Aielli:2009ca, Auger2012, EAStop, abbasi2015}. In addition, 
      the high energy models (QGSJET~II.4, QGSJET01, SIBYLL 2.3, EPOS-LHC)  cross
      section predictions are shown.}
\end{center} 
\vspace{1cm} 
\end{figure}

The final proton-air cross section measured by the Telescope Array detector at an average energy of $10^{18.45}$ eV using the $K$-Factor method and including the statistical and systematic checks is $\sigma^{\mathrm{inel}}_{\mathrm{p-air}} = 520.1 \pm 35.8$ [Stat.] $^{+25}_{-40}$[Sys.]~mb.  This result is shown in Figure~\ref{fig:pair} and is compared to other experimental measurements~\cite{Siohan:1978zk, FE1987, Honda1992, Mielke:1994un, Belov2006, Aielli:2009ca, Auger2012, EAStop, abbasi2015} and current high energy model predictions. Note here that the current proton-air cross section result including the error fluctuations is  consistent with the high energy models tuned to the LHC (QGSJET~II.4~\cite{Ostapchenko:2007qb,Ostapchenko:2010vb}, SIBYLL2.3~\cite{sibyll,Ahn:2009wx}, and EPOS-LHC~\cite{Pierog:2013ria}) shown in Figure~\ref{fig:pair}.

%520-40

\subsection{Proton-Proton Cross Section}

The analysis to convert from the inelastic proton-air cross section to proton-proton cross section consists of two parts. 

\begin{figure}[ht]
\begin{center}\vspace{1cm}
\includegraphics[width=1.0\linewidth]{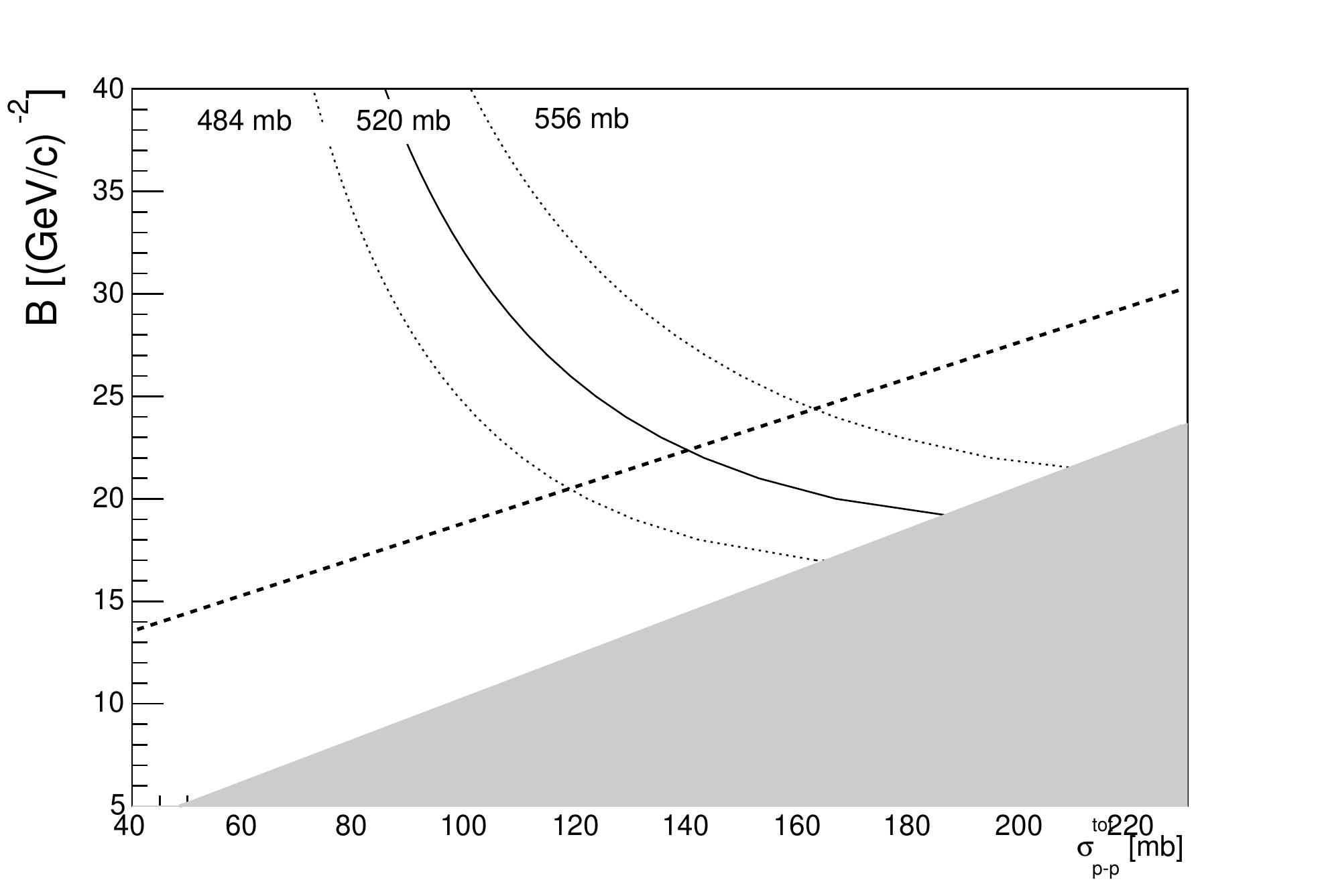}
\vspace{1.0cm}
\caption{\label{fig:glaub} The elastic slope $B$ in
      ((GeV/c)$^{-2}$) vs. $\sigma_{\rm p-p}^{\rm total}$ in mb. The solid 
      line is the allowed $\sigma_{\rm p-p}^{\rm total}$  values from the 
       $\sigma_{\rm p-air}^{\rm inel}$ and its statistical errors reported in this work. 
      The dashed line is the BHS fit prediction~\cite{BHS}. While the gray shaded area is the unitarity constraint.}
\end{center} 
\vspace{1cm} 
\end{figure}

The first part is done by converting the measured inelastic proton-air cross section to the possible allowed values of the proton-proton cross. The conversion is obtained using the Glauber formalism~\cite{Glauber70} which gives $\sigma^{\mathrm{inel}}_{\mathrm{p-air}} $ as a function of $\sigma_{\mathrm{pp}}^{\mathrm{tot}}$ and $B$, where $B$ is the forward scattering elastic slope. The three curved lines in Figure~\ref{fig:glaub} show the TA measurement of $\sigma^{\mathrm{inel}}_{\mathrm{p-air}} $ and its statistical uncertainties allowed region in the ($\sigma_{\mathrm{pp}}^{\mathrm{tot}}$-$B$) plane.

The second part is done by constraining the relation between $\sigma_{\mathrm{pp}}^{\mathrm{tot}}$ and $B$ using a theoretical model.
The model used in this work is (Block, Halzen, and Stanev (BHS))~\cite{Block05}, shown as the dashed line in Figure~\ref{fig:glaub}. The intersection of the $\sigma^{\mathrm{inel}}_{\mathrm{p-air}} $ allowed region and the theoretical constraint (BHS model) gives us $\sigma_{\mathrm{pp}}^{\mathrm{tot}}$ and $B$ values.
 Note that the BHS model can be replaced with other models or predictions to solve for the $\sigma_{\mathrm{pp}}^{\mathrm{tot}}$. Note the BHS model is consistent with the unitarity constraint while describing the pp and $\bar{\mathrm{p}}$p cross section data from the Tevatron  well~\cite{DiasDeDeus:1987bf, Buras:1973km}.

\begin{figure}[ht]
\begin{center}\vspace{1cm}
\includegraphics[width=1.0\linewidth]{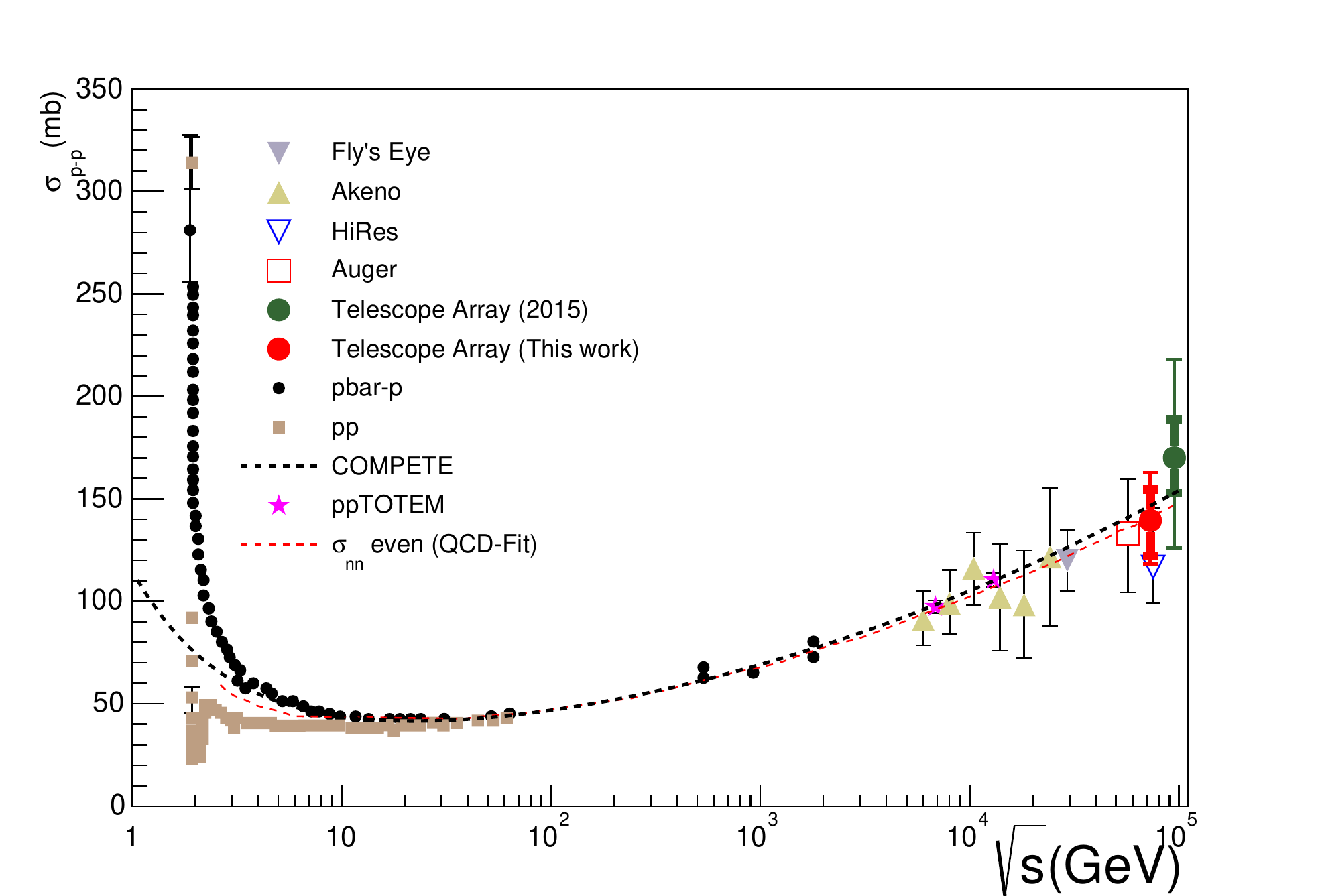}
\vspace{1.0cm}
\caption{\label{fig:pp}  A compilation of the  proton-proton cross section vs. the 
      center of mass energy result of this work, including the statistical (thin) and systematics (thick) error bars, in addition to previous work by cosmic rays detectors~\cite{FE1987,Honda1992,Belov2006,Auger2012,abbasi2015} in addition to, the recent result from LHC by TOTEM at $\sqrt{s}=$ 7 and 13 TeV~\cite{totem7,totem13}. The dashed red curve is the BHS fit~\cite{Block05} and the dashed black curve is the fit by the COMPETE collaboration~\cite{PhysRevLett.89.201801}. This plot is adapted and modified from ~\citep{Block05}.}
\end{center} 
\vspace{1cm} 
\end{figure}

The proton-proton cross section in this work is found to be $\sigma^{\mathrm{tot}}_{\mathrm{pp}} = 139.4 ^{+23.4}_{-21.3}$ [Stat.]$ ^{+15.0}_{-24.0}$[Sys.]~mb. This result is shown in Figure~\ref{fig:pp} in comparison to previously reported values by UHECR experiments~\cite{FE1987,Honda1992,Belov2006,Auger2012,abbasi2015}. The recent result from LHC by TOTEM at $\sqrt{s}=$ 7 and 13 TeV~\cite{totem7,totem13} is also shown, in addition to the BHS fit~\cite{Block05}. The best fit of the proton-proton total cross section data by the COMPETE collaboration is also added~\cite{Compete}. For further details concerning the proton-air to proton-proton procedure refer to \cite{abbasi2015}.

%%%%%%%%%%%%%%%%%%%%%%%%%%%%%%%%%%%%%%%%%%%%%%
%\input{conclusion}

\section{Summary and Conclusion}\label{sec:summary}

Telescope Array has measured the inelastic proton-air cross section of
ultra high energy cosmic rays at $\sqrt{s} = 73$~TeV. This measurement is performed for energies that are not
accessible to accelerator experiments, therefore provides an
important and unique test of standard model predictions about the
fundamental nature of matter.

The Telescope Array utilizes a large array of surface detectors and
fluorescence telescopes to record the atmospheric depth of maximum
size of air showers initiated by inelastic collisions of ultra high
energy cosmic rays and air molecules in the upper atmosphere. By
combining the geometric and timing information of SDs and the Black
Rock Mesa and Long Ridge FDs that observe a hybrid event \xm{} can be
determined with a good precision of $\sim 20$~g/cm$^2$. UHECR \xm{}
distributions are related to the interaction length of cosmic rays in
the atmosphere, which in turn depends on the tail of \xm{} distributions is populated with the deepest
penetrating events, predominantly proton initiated events, the slope
of which is related to the interaction length by a constant,
$K$. Using Monte Carlo simulations $K$ can be evaluated using Monte
Carlo provides access to the depth of first interaction and \xm{} for
each event, allowing a direct determination of $K$. Once $K$ is known
the inelastic proton-air cross section can be determined using
equation~\ref{eq:lambda_m_2}. Using nearly nine years of hybrid data,
TA measures $\sigma^{\mathrm{inel}}_{\mathrm{p-air}} = 520.1 \pm 35.8$
[Stat.] $^{+25}_{-40}$[Sys.]~mb for $\sqrt{s} = 73$~TeV. Using
Glauber theory and the Block, Halzen, Stanev model The total
proton-proton cross section is determined from
$\sigma^{\mathrm{inel}}_{\mathrm{p-air}} $ to be $\sigma^{\mathrm{tot}}_{\mathrm{pp}} = 139.4 ^{+23.4}_{-21.3}$ [Stat.]$ ^{+15.0}_{-24.0}$[Sys.]~mb.

It is interesting to note that ultra high energy cosmic ray model
prediction of the proton-air cross section have converged closer than
was the case prior to tuning to LHC data. This is shown in
the $K$ value converging from 7$\%$ down to 3$\%$. Most importantly,
this is also found to be consistent with results for ultra high energy
cosmic ray experiments including this work. The data from the high
energy models and ultra high energy cosmic ray experiments continue to
show a rising cross section with energy. 

Future cross section results, using TA$\times$4~\cite{2019EPJWC.21006001K} will allow
us to report on the proton air cross section with greater statistical power. Moreover, including data from
the Telescope Array Lower Extension~\cite{Abbasi:2018xsn} would allow the measurement from 10$^{17}$$-$10$^{19}$ eV
with high statistical power and at several energy intervals. 
This would allow us to make
a statement on the functional form of the cross section energy dependence.

%%%%%%%%%%%%%%%%%%%%%%%%%%%%%%%%%%%%%%%%%%%%%%
%\input{acknowledge}

%%%%%%%%%%%%%%%%%%%%%%%%%%%
\section{Acknowledgements}
The Telescope Array experiment is supported by the Japan Society for
the Promotion of Science(JSPS) through
Grants-in-Aid
for Priority Area
%"Highest Energy Cosmic Rays"
431,
for Specially Promoted Research
%``Extreme Phenomena in the Universe Explored by Highest Energy Cosmic Rays''
%Grant Number
JP21000002,
%Grant-in-Aid
for Scientific  Research (S)
%"Quest for the unified picture of the explosion mechanism of supernovae and the central engine of gamma-ray bursts"
%Grant Number
JP19104006,
%Grant-in-Aid
for Specially Promoted Research
%"Extended Telescope Array Experiment - Nearby Extreme Universe Elucidated by Highest-energy Cosmic Rays"
%Grant Number
JP15H05693,
%Grant-in-Aid
for Scientific  Research (S)
%"Study of the ultra high energy cosmic ray source evolution by detailed measurement of cosmic rays in the wide energy range"
%Grant Number
JP15H05741, for Science Research (A) JP18H03705,
%Grant-in-Aid
for Young Scientists (A)
%"hoge hoge"
%Grant Number
JPH26707011,
and for Fostering Joint International Research (B)
%"Search for Ultra-High Energy Cosmic Ray origin using the extended Telescope Array experiment"
%Grant Number
JP19KK0074,
by the joint research program of the Institute for Cosmic Ray Research (ICRR), The University of Tokyo;
by the U.S. National Science
Foundation awards PHY-0601915,
PHY-1404495, PHY-1404502, and PHY-1607727;
by the National Research Foundation of Korea
% \linebreak
(2016R1A2B4014967, 2016R1A5A1013277, 2017K1A4A3015188, 2017R1A2A1A05071429) ;
%\linebreak
by the Russian Academy of
Sciences, RFBR grant 20-02-00625a (INR), IISN project No. 4.4502.13, and Belgian Science Policy under IUAP VII/37 (ULB). The foundations of Dr. Ezekiel R. and Edna Wattis Dumke, Willard L. Eccles, and George S. and Dolores Dor\'e Eccles all helped with generous donations. The State of Utah supported the project through its Economic Development Board, and the University of Utah through the Office of the Vice President for Research. The experimental site became available through the cooperation of the Utah School and Institutional Trust Lands Administration (SITLA), U.S. Bureau of Land Management (BLM), and the U.S. Air Force. We appreciate the assistance of the State of Utah and Fillmore offices of the BLM in crafting the Plan of Development for the site.  Patrick Shea assisted the collaboration with valuable advice  on a variety of topics. The people and the officials of Millard County, Utah have been a source of steadfast and warm support for our work which we greatly appreciate. We are indebted to the Millard County Road Department for their efforts to maintain and clear the roads which get us to our sites. We gratefully acknowledge the contribution from the technical staffs of our home institutions. An allocation of computer time from the Center for High Performance Computing at the University of Utah is gratefully acknowledged.
%%%%%%%%%%%%%%%%%%%%%%%%%%%%%%%%%%%%%%%%%%%%%%%%%%%%%%%%%%%%%%%%%%%%%%%%%%%%%%%%xss

%%%%%%%%%%%%%%%%%%%%%%%%%%%%%%%%%%%%%%%%%%%%%%
%\bibliography{paper}{}


\end{document}